\documentclass[article,aps,nofootinbib,showpacs,amsfonts,epsf]{revtex4}

\input{epsf.tex}

\newcommand{\E}{{\cal{E}}}

\newcommand{\s}{\sigma}

\renewcommand{\d}{{\rm d}}
\renewcommand{\a}{\alpha}
\renewcommand{\k}{\kappa}
\newcommand{\dfrac}[2]{\displaystyle\frac{#1}{#2}}
\newcommand{\be}{\begin{equation}}
\newcommand{\ee}{\end{equation}}
\newcommand{\bea}{\begin{eqnarray}}
\newcommand{\eea}{\end{eqnarray}}
\newcommand{\ba}{\begin{array}}
\newcommand{\ea}{\end{array}}

\def\PRD{Phys. Rev. D}
\def\PR{Phys. Rev.}
\def\PRL{Phys. Rev. Lett.}
\def\PTP{Prog. Theor. Phys.}

\def\JMP{J. Math. Phys.}

\def\PLA{Phys. Lett. A}

\begin{document}
\draft

\title{On a Simple Representation of the Kinnersley-Chitre Metric}

\author{V.~S.~Manko$^\dag$ and E.~Ruiz$\,^\ddag$}
\address{$^\dag$Departamento de F\'\i sica, Centro de Investigaci\'on y de
Estudios Avanzados del IPN, A.P. 14-740, 07000 M\'exico D.F.,
Mexico\\$^\ddag$Instituto Universitario de F\'{i}sica
Fundamental y Matem\'aticas, Universidad de Salamanca, 37008 Salamanca, Spain}

\begin{abstract}
A concise form of the Kinnersley-Chitre five-parameter metric for a spinning mass is obtained by exploiting a remarkable similarity between the metric's factor structure and the analogous structure of the Tomimatsu-Sato solutions with even distortion parameter $\delta$. The corresponding general subfamily of asymptotically flat spacetimes containing four arbitrary real parameters is considered, and all configurations describing two extreme Kerr sources separated by a strut are identified and briefly discussed. \end{abstract}


\maketitle

%\twocolumn

\section{Introduction}

The Kinnersley-Chitre (K-C) five-parameter vacuum solution \cite{KCh} describing the gravitational field of a stationary axisymmetric mass distribution is historically one of the first most interesting and significant applications of the modern solution generating techniques to Einstein's theory. It generalizes the well-known Tomimatsu-Sato (T-S) $\delta=2$ solution for a spinning mass~\cite{TSa} and represents the extreme limit of the double-Kerr solution of Kramer and Neugebauer~\cite{KNe}. In the case of zero NUT parameter, the metric functions defined by the Ernst potential~\cite{Ern} of the paper~\cite{KCh} were obtained by Yamazaki~\cite{Yam} who employed for them a representation analogous to the one he earlier discovered~\cite{Yam2} for the T-S metrics with integer distortion parameter $\delta$~\cite{TSa2}. Although the Yamazaki's result is very useful and important as it permits one to work out the whole set of metrical fields for any\footnote{Some problems in the use of the results of Ref.~[5] for elaborating particular ($q=0$) cases were mentioned in the paper \cite{MRRS}.} particular asymptotically flat specialization of the K-C solution, still it can be observed that the function $\omega$ was given in Ref.~[5] in a by far more complicated form than the other two metric functions, $f$ and $\gamma$, which suggests the existence of a simpler representation of the K-C metric.

In the present paper we will show that a concise form of the {\it general} five-parameter K-C metric is possible and it can be obtained by extending the results of Perj\'es' work \cite{Per}, where the factor structure of T-S solutions was studied, to spacetimes with a non-zero NUT parameter. Thus we shall be able, on the one hand, to write down the general K-C metric with the aid of only four basic polynomials and, on the other hand, to reconsider separately the general asymptotically flat subclass of the K-C metric for rewriting it in a much simpler form than in Ref.~[5]. Within the latter subclass we shall point out and briefly discuss an interesting special case representing two extreme Kerr black holes separated by a massless strut which has not yet been considered in the literature except for a very particular configuration of identical counterrotating extreme black holes \cite{MRRS}.

Our paper is organized as follows. In Section 2 we introduce the K-C solution via the Ernst complex potential and consider its behavior on the symmetry axis. In Section 3 the metric functions of K-C solution will be given in the Perj\'es-like representation, and finally in Section 4 we shall consider the asymptotically flat subclass of the general solution, with an emphasis on a binary system for two extreme Kerr black holes kept apart by a strut.

\section{The Ernst potential and axis data of K-C solution}

As is well known \cite{Ern}, the stationary axially symmetric vacuum problem reduces to solving the Ernst equation
\be
(\E+\bar\E)(\E_{,\rho,\rho}+\rho^{-1}\E_{,\rho}+\E_{,z,z})= 2(\E_{,\rho}^2+\E_{,z}^2) \label{E_eq} \ee
for a complex potential $\E$, where a bar over a symbol means complex conjugation, $\rho$ and $z$ are the Weyl-Papapetrou cylindrical coordinates, and comma denotes partial differentiation. The relation of $\E$ to the metric functions $f$, $\omega$ and $\gamma$ from Papapetrou's line element \cite{Pap}
\be
d s^2=f^{-1}[e^{2\gamma}(d\rho^2+d z^2)+\rho^2\d\varphi^2]-f(d
t-\omega d\varphi)^2 \label{Papa} \ee
is defined by the equations
\bea
f&=&{\rm Re}(\E), \nonumber\\ \omega_{,\rho}&=&-4\rho(\E+\bar\E)^{-2}{\rm Im}(\E_{,z}), \nonumber\\
\omega_{,z}&=&4\rho(\E+\bar\E)^{-2}{\rm Im}(\E_{,\rho}), \nonumber\\ \gamma_{,\rho}&=&\rho(\E+\bar\E)^{-2}(\E_{,\rho}\bar\E_{,\rho} -\E_{,z}\bar\E_{,z}), \nonumber\\ \gamma_{,z}&=&2\rho(\E+\bar\E)^{-2}{\rm Re}(\E_{,\rho}\bar\E_{,z}), \label{mf_eq} \eea
which involve the real (Re) and imaginary (Im) parts of the potential $\E$.

The K-C solution was originally given in terms of another Ernst potential $\xi$ related to $\E$ by the formula
\be
\xi=\frac{1-\E}{1+\E} \quad\Longleftrightarrow\quad \E=\frac{1-\xi}{1+\xi}, \label{ksi} \ee
and its explicit form is the following:\cite{KCh}
\bea
\xi&=&B/A, \nonumber\\
A&=&p^2(x^4-1)+(\a^2-\beta^2)(x^2-y^2)^2+q^2(y^4-1) -2ipqxy(x^2-y^2) \nonumber\\
&-&2i\a(x^2+y^2-2x^2y^2)-2i\beta xy(x^2+y^2-2), \nonumber\\
B&=&2(P-iQ)[px(x^2-1)+iqy(y^2-1)-i(p\a+iq\beta)x(x^2-y^2) \nonumber\\ &+&i(p\beta+iq\a)y(x^2-y^2)], \label{ksi_KC} \eea
where we have used the original K-C notation for the parameters $p$, $q$, $\a$, $\beta$, but changed exp($-i\gamma$) of the paper \cite{KCh} to our $P-iQ$. Note that $\a$ and $\beta$ can take on arbitrary real values, while $p$, $q$, $P$ and $Q$ represent only two arbitrary real constants, being subjected to the restrictions
\be
p^2+q^2=1, \quad P^2+Q^2=1. \label{pq} \ee
Equations (\ref{ksi_KC}) are written in prolate spheroidal coordinates $x$ and $y$ defined as
\be
x=\frac{1}{2\k}(r_++r_-), \quad y=\frac{1}{2\k}(r_+-r_-), \quad r_\pm=[\rho^2+(z\pm\k)^2]^{1/2}, \label{xyrz} \ee
where $\k$ is the fifth independent parameter of K-C solution. The inverse transformation is
\be
\rho=\k(x^2-1)^{1/2}(1-y^2)^{1/2}, \quad z=\k xy. \label{rzxy} \ee

Apparently, the Ernst potential $\E=(A-B)/(A+B)$, with $A$ and $B$ given by (\ref{ksi_KC}), satisfies Eq.~(\ref{E_eq}). It is worth noting that $\E$ on the upper part of the symmetry axis ($y=1$, $x=z/\kappa$) takes the form
\bea
&&\E(\rho=0,z)=e_+/e_-, \nonumber\\
&&e_\pm=(p^2+\a^2-\beta^2)z^2-2\k[\pm(P-iQ)(p+q\beta-ip\a)+i(pq+\beta)]z \nonumber\\ &&\hspace{2cm}+\k^2(p^2-\a^2+\beta^2+2i\a)\pm2\k^2(P-iQ)(q\a-ip\beta), \label{E_axis} \eea
and the knowledge of the above axis data is sufficient for reconstructing $\E$ in the whole space and obtaining the corresponding metric functions with the aid of Sibgatullin's integral method \cite{Sib,MSi}. In Appendix we give the expressions of the metric coefficients $f$, $\omega$ and $\gamma$ for the K-C solution in terms of the parameter $\k$ and two poles of the function $e_-$, which are essential for obtaining the main results of Section 3.

Using (\ref{E_axis}), it is easy to find the total mass $M$ and total angular momentum $J$ of K-C solution by employing the Fodor-Hoenselaers-Perj\'es procedure \cite{FHP} for the calculation of Geroch-Hansen multipole moments \cite{Ger,Han}. The result is\footnote{The form of $J$ is given for the value $C=0$ in Eqs.~(\ref{mf_gen}).}
\be
M=\frac{2\k(pP-pQ\a+qP\beta)}{p^2+\a^2-\beta^2}, \quad J=\frac{M[(pq+\beta)M+\k(qQ\a+pP\beta)]}{pP-pQ\a+qP\beta}, \label{MJ} \ee
and it is likely to add to these expressions a formula for the monopole angular momentum moment $J_0$, namely,
\be
J_0=-\frac{2\k(pQ+pP\a+qQ\beta)}{p^2+\a^2-\beta^2}, \label{J0} \ee
because the condition $J_0=0$ determines the asymptotically flat subclass of K-C solution. In the paper \cite{Yam} the asymptotic flatness was achieved by solving the latter condition for $P$ and $Q$:
\be
\frac{Q}{P}=-\frac{p\a}{p+q\beta}, \ee or \be P=\frac{p+q\beta}{\sqrt{(p+q\beta)^2+p^2\a^2}}, \quad Q=-\frac{p\a}{\sqrt{(p+q\beta)^2+p^2\a^2}}, \label{J0_Yam} \ee
thus introducing irrational quantities  into the formulas defining the asymptotically flat family of K-C spacetimes. In Section 4, however, the condition $J_0=0$ will be solved for $\a$, with the idea to explore more efficiently the factorization properties of K-C solution.

\section{Metric functions}

The metric functions $f$, $\gamma$, $\omega$ of K-C solution which are given below have been worked out with the aid of formulas (\ref{A1})-(\ref{A3}) of Appendix, and they are written, following Perj\'es \cite{Per}, in terms of four polynomials $\mu$ ($=\rho$ in Perj\'es' notation), $\sigma$, $\pi$ and $\tau$:
\bea f&=&\frac{N}{D}, \quad e^{2\gamma}=\frac{N}{K_0^2(x^2-y^2)^4}, \quad \omega=2J_0(y+C) +\frac{\k(y^2-1)F}{N}, \nonumber\\
N&=&\mu^2+(x^2-1)(y^2-1)\s^2, \nonumber\\ D&=&N+\mu\pi-(y^2-1)\s\tau, \nonumber\\ F&=&(x^2-1)\s\pi+\mu\tau, \nonumber\\ \mu&=&p^2(x^2-1)^2+q^2(y^2-1)^2+(\a^2-\beta^2)(x^2-y^2)^2, \nonumber\\ \s&=&2[pq(x^2-y^2)+\beta(x^2+y^2)-2\a xy], \nonumber\\ \pi&=&(4/K_0)\{K_0 [pPx(x^2+1)+2x^2+qQy(y^2+1)] \nonumber\\ &+&2(pQ+pP\a+qQ\beta)[pqy(x^2-y^2)+\beta y(x^2+y^2)-2\a xy^2] \nonumber\\ &-&K_0(x^2-y^2)[(pQ\a-qP\beta)x+(qP\a-pQ\beta)y] \nonumber\\ &-&2(q^2\a^2+p^2\beta^2)(x^2-y^2)+4(pq+\beta)x(\beta x-\a y)\}, \nonumber\\ \tau&=&(4/K_0)\{K_0 x [(qQ\a+pP\beta)(x^2-y^2)+qP(y^2-1)] \nonumber\\ &+&(pQ+pP\a+qQ\beta)y[(p^2-\a^2+\beta^2)(x^2-y^2)+y^2-1] \nonumber\\ &-&pQK_0y(x^2-1)-2p(q\a^2-q\beta^2-p\beta)(x^2-y^2) \nonumber\\ &+&(pq+\beta)(y^2-1)\}, \nonumber\\ K_0&=&p^2+\a^2-\beta^2, \label{mf_gen} \eea
where $J_0$ is defined by Eq.~(\ref{J0}). The integration constant $C$ has been introduced in analogy with our paper \cite{MRu} on the physical interpretation of the NUT spacetime \cite{NTU}: for non-vanishing $J_0$, the values $C=\pm1$ define the cases with one semi-infinite singularity along the symmetry axis, while the choice $C=0$ leads to the only asymptotically non-flat case with a finite angular momentum.

A crucial step for obtaining Eqs.~(\ref{mf_gen}) was to find a correct decomposition of the metric function $\omega$ which could permit us to extend Perj\'es' results he discovered for the asymptotically flat T-S solutions, to a more general case characterized by a non-vanishing NUT parameter $J_0$. In this respect, singling out the asymptotically vanishing part of $\omega$ was a key point for eventually obtaining our representation of the K-C metric, almost identical to that of the T-S spacetimes with even distortion parameter $\delta$. It is worth noting that Yamazaki's paper \cite{Yam} helped us to guess the form of polynomials $\mu$ and $\sigma$, while the remaining functions $\pi$ and $\tau$ were found from the cumbersome ``interim'' expressions for $f$ and $\omega$. In spite of its unphysical character, the NUT parameter $J_0$ is likely to be included in the above formulas, first of all, for historical reasons requiring the description of the most general, five-parameter K-C spacetime and any particular specialization of the latter; furthermore, one may also think about possible applications of the general K-C solution in electrostatics or magnetostatics (for instance, by using Bonnor's theorem \cite{Bon}) where the NUT parameter would play the role of an electric (or magnetic) charge.

Eqs.~(\ref{ksi_KC}), (\ref{pq}) and (\ref{mf_gen}) permit one to elaborate any specific case of the K-C solution.

\section{The asymptotically flat family}

The condition $J_0=0$ defines the asymptotically flat four-parameter subclass of K-C spacetimes, most interesting from the physical point of view, and it follows from (\ref{J0}) that there are several ways to satisfy the asymptotic flatness condition. Formulas (\ref{J0}) display Yamazaki's way of solving equation $J_0=0$ that was utilized in the paper \cite{Yam}, and such choice leads to the irrational quantities in Yamazaki's formulas. For that reason, in what follows we shall explore another possibility to achieve vanishing $J_0$, namely,
\be
\a=-\frac{Q(p+q\beta)}{pP}, \label{J0_MR} \ee
which permits one to avoid irrational expressions and, as a result, is advantageous for the analysis of the axis behavior of the metric function $\omega$.

The substitution of (\ref{J0_MR}) into Eqs.~(\ref{ksi_KC}) and (\ref{mf_gen}) then gives us the form of the Ernst potential and corresponding metric functions of the asymptotically flat K-C solution:
\bea
\xi&=&\frac{B}{A}, \quad
f=\frac{N}{D}, \quad e^{2\gamma}=\frac{N}{K_0^2(x^2-y^2)^4}, \quad \omega=\frac{\k(y^2-1)F}{N}, \nonumber\\
A&=&p^2P^2\{p^2(x^4-1)+q^2(y^4-1)-2ixy[pq(x^2-y^2)+\beta(x^2+y^2-2)]\} \nonumber\\ &+&[Q^2(p+q\beta)^2-p^2P^2\beta^2](x^2-y^2)^2
+2ipPQ(p+q\beta)(x^2+y^2-2x^2y^2), \nonumber\\
B&=&2pP(P-iQ)\{(x^2-y^2)[pP\beta(qx+ipy)+Q(p+q\beta)(qy+ipx)] \nonumber\\ &+&pP[px(x^2-1)+iqy(y^2-1)]\}, \nonumber\\
N&=&\mu^2+(x^2-1)(y^2-1)\s^2, \nonumber\\ D&=&N+\mu\pi-(y^2-1)\s\tau, \nonumber\\ F&=&(x^2-1)\s\pi+\mu\tau, \nonumber\\
\mu&=&p^2P^2[p^2(x^2-1)^2+q^2(y^2-1)^2]+[Q^2(p+q\beta)^2-p^2P^2\beta^2](x^2-y^2)^2, \nonumber\\ \s&=&2pP\{pP[pq(x^2-y^2)+\beta(x^2+y^2)]+2Q(p+q\beta)xy\}, \nonumber\\ \pi&=&(4pP/K_0)\{K_0 pP[pPx(x^2+1)+2x^2+qQy(y^2+1)] \nonumber\\ &+&K_0(x^2-y^2)[(p^2Q^2+pq\beta)x+PQ(pq+\beta)y] \nonumber\\ &-&2pP[q^2Q^2(p+q\beta)^2+p^4P^2\beta^2](x^2-y^2) \nonumber\\ &+&4p^2P^2(pq+\beta)x[pP\beta x+Q(p+q\beta)y)]\}, \nonumber\\ \tau&=&(4pP/K_0)\{K_0 x [pqP^2(x^2-1)-(pq-p^2\beta+Q^2\beta)(x^2-y^2)] \nonumber\\ &-&2p^2P(p+q\beta)(pqQ^2-p^2P^2\beta+q^2Q^2\beta)(x^2-y^2) \nonumber\\ &-&p^2PQK_0y(x^2-1)+p^3P^3(pq+\beta)(y^2-1)\}, \nonumber\\ K_0&=&p^2P^2(p^2-\beta^2)+Q^2(p+q\beta)^2, \label{mf_af_gen} \eea
while the expressions (\ref{MJ}) for the total mass and total angular momentum rewrite as
\bea
&&M=\frac{2\kappa p^2P(p+q\beta)}{p^2P^2(p^2-\beta^2)+Q^2(p+q\beta)^2}, \nonumber\\ &&J=M\left[\frac{P(pq+\beta)M}{p+q\beta}-\frac{\kappa}{p} \left(qQ^2-\frac{p^2P^2\beta} {p+q\beta}\right)\right]. \label{MJ_af} \eea

The known stationary limits of the metric (\ref{mf_af_gen}) are the T-S $\delta=2$ solution for a spinning mass \cite{TSa} ($\beta=Q=0$, $P=1$), Tomimatsu's configuration for two balancing extreme Kerr sources \cite{Tom,Hoe} ($p=q=-P=Q=1/\sqrt{2}$, $\beta=-(1+l)(2l)^{-1}$), an equilibrium configuration of extreme sources due to Dietz and Hoenselaers \cite{DHo} (the particular choice of parameters leading to this limit is given later on in the text, see Eq.~(\ref{b_DH})). Moreover, by choosing $\beta=q=0$, $p=1$, one arrives at the solution for a pair of identical counterrotating extreme black holes separated by a conical singularity which has recently been considered in the paper \cite{MRRS}, and the particular case $P=1$, $Q=0$, also arising as a vacuum specialization of the solution \cite{MMS}, was shown to be appropriate for modeling the exterior field of rapidly rotating neutron stars \cite{SCa,BSt}.

Remarkably, a very simple form of Eqs.~(\ref{mf_af_gen}) permits us to solve the general problem of two extreme Kerr black holes separated by a massless strut. Mention that in the non-extreme case the analogous problem has not yet been solved due to complexity of the corresponding algebraic equations involved in its resolution (see, e.g., Ref.~[21], p.~349), so that one might expect that the extreme case could represent even more technical difficulties than the non-extreme one. Fortunately, this is not so, and in order to single the desired two-black hole subclass out of the asymptotically flat K-C solution (\ref{mf_af_gen}) it is only necessary to solve the axis condition which consists in vanishing of the metric function $\omega$ on the part of the symmetry axis separating two black holes ($\rho=0$, $-\kappa<z<\kappa$); the factotization of $\omega$ into smaller factors then ensures the resolution of this condition in the general case. Indeed, the axis condition yields
\bea
&&\omega(x=1)=0 \quad \Longrightarrow \quad \omega_1\omega_2=0, \nonumber\\
&&\omega_1=(p^2-Q^2)\beta-pq(pP+Q^2), \nonumber\\
&&\omega_2=(p^2-Q^2)\beta^2-pq(1+pP+Q^2)\beta-p^2(1+pP), \label{w_axis} \eea
thus giving rise to two subfamilies of solutions which we consider below.

\subsection*{A. The subfamily defined by $\omega_1=0$.}

In this case we have a linear equation for $\beta$ which gives
\be
\beta=\frac{pq(pP+Q^2)}{p^2-Q^2}, \label{b_1} \ee
and the simplest choice $q=0$ in (\ref{b_1}) leads to the solution \cite{MRRS} for two identical counter-rotating extreme Kerr black holes. Accounting for (\ref{b_1}), formulas (\ref{MJ_af}) for the total mass and total angular momentum assume the form
\be
M=\frac{2\kappa(pP+q^2)}{p^2-q^2}, \quad J=\frac{2\kappa^2q[(1+2pP)^2-(p+P)^2]}{p(p^2-q^2)^2}. \label{MJ_s1} \ee
The individual Komar \cite{Kom} masses and angular momenta of the constituents can be worked out with the aid of the results of papers \cite{Tom,DHo}. Thus, for instance, by choosing
\be
\kappa=2, \quad p=4/5, \quad q=3/5, \quad P=Q=1/\sqrt{2}, \label{ex1} \ee
we obtain
\bea
&&M_1=5\sqrt{2}+6, \quad M_2=(5\sqrt{2}-6)/7, \nonumber\\ &&J_1=25(5\sqrt{2}+7)/2, \quad J_2=-25(5\sqrt{2}+1)/98, \label{MiJi_s1} \eea
where the subscript 1 denotes the Komar mass and angular momentum of the upper constituent located at the point $z=\kappa$ of the $z$-axis, while the Komar quantities of the lower constituent located at the point $\rho=0$, $z=-\kappa$, are labeled with subscript~2. Note that in the above example the angular momenta $J_1$ and $J_2$ have opposite signs, whence we tentatively conclude that the whole subfamily (\ref{b_1}) describes two counter-rotating extreme black holes separated by a strut.

\subsection*{B. The subfamily defined by $\omega_2=0$.}

Instead of a linear equation that we had in the previous case, we now have a quadratic equation which  readily gives
\be
\beta=\frac{p[q(1+pP+Q^2)\pm P\Delta]}{2(p^2-Q^2)}, \quad \Delta=[4p^2(1+pP)+q^2(p+P)^2]^{1/2}, \label{b_2} \ee
the corresponding expressions for the total mass and total angular momentum being
\bea
&&M=\frac{\kappa[\pm q\Delta-p(1+p^2)-q^2P]}{p(p^2-q^2)}, \nonumber\\
&&J=\frac{\kappa M}{2p(p^2-q^2)}[\pm\Delta(2p^2P-2p-P)+2q(1+p^2+pP) \nonumber \\ &&\hspace{3cm}-qP(p^2-q^2)(p-P)]]. \label{MJ_s2} \eea

It is easy to see that in Eqs.~(\ref{b_2}), (\ref{MJ_s2}) one can always restrict himself to the upper, `plus' sign because the case defined by the `minus' sign is obtainable from the former by changing $q$ to $-q$ and $J$ to $-J$.

For the calculation of individual Komar masses and angular momenta of the extreme constituents one can again make use of the general formulas of papers \cite{Tom,DHo}. Below we give an example of a system involving positive masses of both components:
\be
\kappa=2, \quad p=3/5, \quad q=-4/5, \quad P=Q=1/\sqrt{2}, \label{ex2} \ee
leading to
\bea
&&M_1=\left(3606+3830\sqrt{2}+\sqrt{40963436+26976210\sqrt{2}}\right)/861\simeq 20.81, \nonumber\\ &&M_2\simeq 11.18, \quad J_1\simeq 606.587, \quad J_2\simeq 405.821 \label{MiJi_s2} \eea
(the approximate values are given up to three decimal places), and in (\ref{MiJi_s2}) we provide an exact value only for $M_1$ because of cumbersome explicit formulas for $M_i$ and $J_i$ in this particular case.

A physically interesting special 2-parameter member of this family is a solution describing two corotating identical extreme Kerr constituents separated by a massless strut which is defined by the following choice of the parameters in (\ref{mf_af_gen}):
\be
Q=0, \quad P=1, \quad \beta=\frac{1}{2p}[\Delta_{\rm S}+q(1+p)], \quad \Delta_{\rm S}=[(1+p)(1+3p^2+pq^2)]^{1/2}. \label{par_es} \ee

Taking into account that the sources are identical and corotating, the individual masses and angular momenta are equal half the respective total quantities, thus yielding
\bea
&&M_1=M_2=\frac{\kappa[q(\Delta_{\rm S}-q)-p(1+p^2)]}{2p(p^2-q^2)}, \nonumber\\
&&J_1=J_2=\frac{\kappa^2[\Delta_{\rm S}(1-p+p^2+4pq^2)-q(1+p)(1+3p+3p^2)+4p^4q]} {2p^2(p^2-q^2)^2}, \label{MJ_es} \eea
whence we obtain an important relation between $M_i$ and $J_i$:
\be
\delta_{\rm S}=\frac{J_1}{M_1^2}=\frac{J_2}{M_2^2}=\frac{1}{2}[\Delta_{\rm S}(2-p)-pq(3-p)]. \label{MJ_rel} \ee

The positive values of $M_i$ which define the black-hole sector of this symmetric configuration correspond to the parameter ranges
\be
-\frac{1}{\sqrt{2}}<p<0, \quad q>0 \qquad \mbox{and} \qquad 0<p<\frac{1}{\sqrt{2}}, \quad q<0, \label{M_pos} \ee
and it is easy to see that, for any $p$ and $q$ satisfying (\ref{M_pos}), the corresponding values of $\delta_{\rm S}$ in (\ref{MJ_rel}) all lie within the interval (1, 2), as illustrated in Figs.~1 and 2. Therefore, the ratios $|J_i|/M_i^2$ in the last example of interacting extreme black holes exceed the value $|J|/M^2=1$ characterizing  a single Kerr black hole \cite{Ker} in the extreme limit, which supports analytically a recent numerical analysis of this configuration carried out by Costa {\it et al} \cite{CHR}. It is worth noting, however, that whereas the values of $|J_i|/M_i^2$ in (\ref{MJ_rel}) cannot be greater than 2, the latter ratio can achieve larger values when the corotating extreme black holes are non-equal: for instance, in the previous example defined by (\ref{ex2}) we had $J_1/M_1^2\simeq 1.401$ and $J_2/M_2^2\simeq 3.247$.

\begin{figure}[htb]
\centerline{\epsfysize=75mm\epsffile{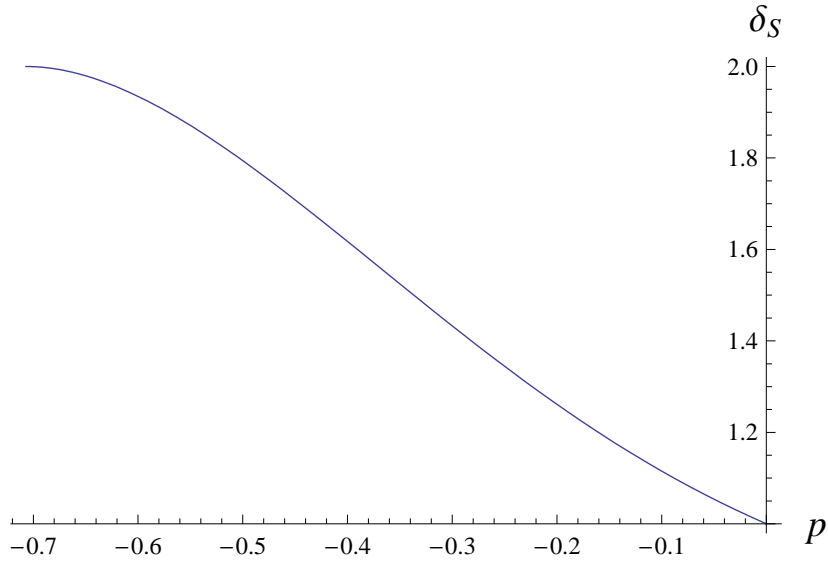}} \caption{Plot of $\delta_{\rm S}$ against $p$ for positive values of the total mass in the case $-1/\sqrt{2}<p<0$, $q>0$.}
\end{figure}

%\newpage

\begin{figure}[htb]
\centerline{\epsfysize=75mm\epsffile{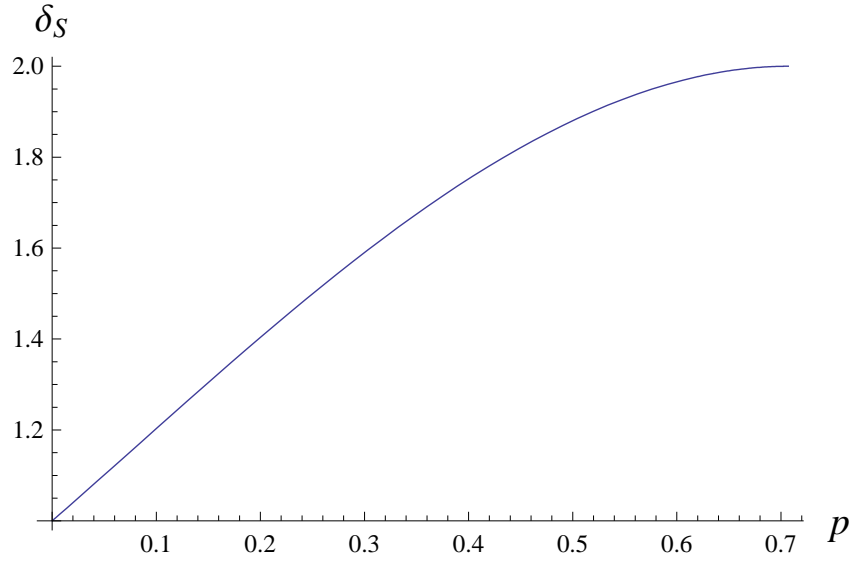}} \caption{Plot of $\delta_{\rm S}$ against $p$ for positive values of the total mass in the case $0<p<1/\sqrt{2}$, $q<0$.}
\end{figure}

We find it instructive to plot the stationary limit surfaces (SLS), defined by $f=0$, for several particular choices of the parameters $p$ and $q$ in the symmetric solution, with the idea to illustrate different physical situations this solution may describe. In Fig.~3(a) the individual SLS of the extreme black holes are disconnected, but these can also merge in one common SLS, as shown in Fig.~3(b), where the SLS are touching each other, and in Fig.~4(a), where the common SLS is already completely formed. It is important to note that in the above three particular configurations with positive Komar masses the ring singularities off the symmetry axis are absent, in contrast to the binary systems of extreme constituents involving a negative mass \cite{Hoe}. This is something expected. On the other hand, a very unexpected result is shown in Fig.~4(b) where two extreme Kerr constituents endowed with positive masses still develop, contrary to the expectations, a massless ring singularity outside the symmetry axis! This interesting phenomenon related to the double-Kerr solution has never been reported before, but it has some similarity with the formation of an analogous singularity in the equilibrium configurations of three Kerr black holes \cite{MRM} and, in our opinion, is a reflection of highly unstable processes which can take place during the formation of a common SLS.

\begin{figure}[htb]
\centerline{\epsfysize=70mm\epsffile{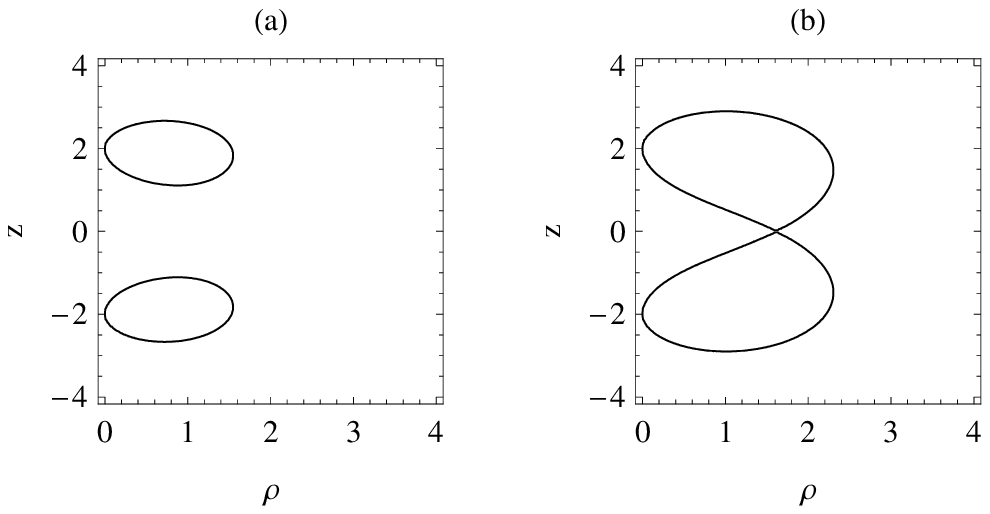}} \caption{The stationary limit surfaces (SLS) of two particular configurations of identical corotating extreme black holes: (a) the case of a SLS consisting of two disconnected parts and corresponding to $p=-0.4$, $q\simeq0.954$; (b) the case of touching SLS, it corresponds to the parameter choice $p=-0.45$, $q\simeq0.893$. No one of these cases develops a massless ring singularity off the symmetry axis.}
\end{figure}

\begin{figure}[htb]
\centerline{\epsfysize=70mm\epsffile{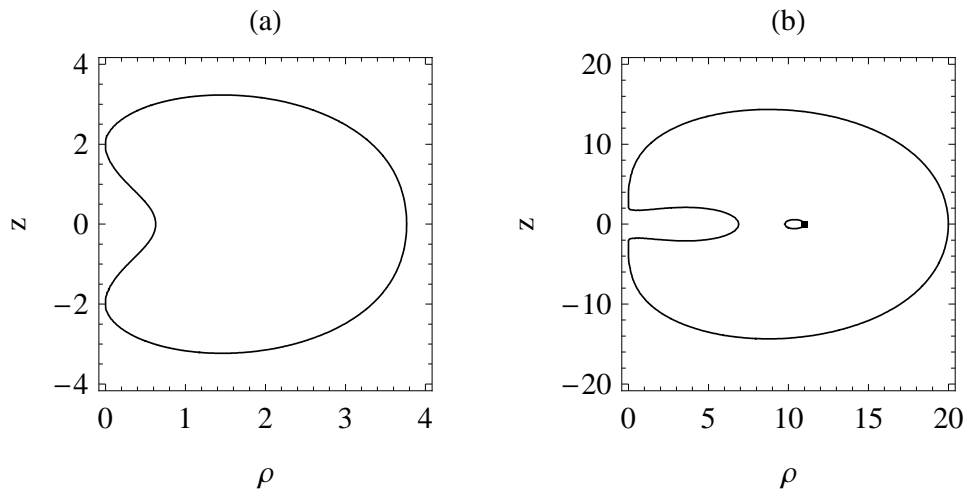}} \caption{Formation of a common SLS by two identical corotating extreme black holes: (a) the parameter choice $p=-0.5$, $q\simeq0.866$ does not lead to the appearance of ring singularities off the symmetry axis; (b) the case $p=0.2$, $q\simeq-0.98$ is characterized by a massless ring singularity located at $z=0$, $\rho\simeq10.987$.}
\end{figure}

\subsection*{C. Three special subfamilies.}

Three more subfamilies of configurations describing a pair of extreme Kerr constituents kept apart by a massless strut arise as special cases corresponding to simultaneous vanishing of the denominator and numerator on the right hand side of Eq.~(\ref{J0_MR}). The first case is defined by the parameter choice
\be
p=0, \quad q=1, \quad P=1, \quad Q=0, \label{par_sc1} \ee
and then the axis condition $\omega(x=1)=0$, after the use of the general formulas (\ref{mf_gen}) of Section 3, gives
\be
\a=\pm\sqrt{\beta^2-\beta}. \label{ab} \ee
For any choice of sign in (\ref{ab}), we arrive at the same negative value of the total mass
\be
M=-2\kappa, \label{M_sc1} \ee
and hence this particular subfamily does not represent much physical interest.

The other two possibilities to satisfy the axis condition follow from the parameter choice
\be
P=0, \quad Q=1, \quad \beta=-p/q, \label{par_sc2} \ee
for which the axis condition yields the equation
\be
(\a-p)(q^2\a^2-pq^2\a-p^2)=0. \label{ac_sc2} \ee

Setting to zero the first factor, we get
\be
\a=p, \label{sc2_1} \ee
and this solution of Eq.~(\ref{ac_sc2}) leads to the expression for the total mass
\be
M=\frac{2\kappa q^2}{p^2-q^2}, \label{M_sc2_1} \ee
which is positive when $p^2>q^2$. Nevertheless, it can be shown that in this case the corresponding individual Komar masses have opposite signs.

The remaining solution of Eq.~(\ref{ac_sc2}), arising from the second factor quadratic in $\a$, is
\be
\a=\frac{p}{2q}(q\pm\sqrt{4+q^2}), \label{sc2_2} \ee
and it admits the configurations of extreme black holes. Indeed, restricting ourselves to the upper sign in (\ref{sc2_2}), we obtain for the total mass the expression
\be
M=\frac{\kappa(q\sqrt{4+q^2}-1-p^2)}{p^2-q^2}, \label{M_sc2_2} \ee
which takes positive values when $q$ belongs to the interval ($-1$,$-1/\sqrt{2}$). Importantly, the corresponding individual masses of the constituents are positive on the same interval too, and a possible typical example is
\be
\kappa=2, \quad p=3/5, \quad q=-4/5,  \label{par_sc2} \ee
leading to
\be
M_1\simeq 18.467, \quad M_2\simeq 3.556, \quad J_1\simeq 366.298, \quad J_2\simeq 122.267, \label{MJ_sc2} \ee
so that the extreme black holes in this particular configuration are corotating.

We end up this section by noting that the balance condition $\gamma(x=1)=0$ which, if satisfied, leads to the disappearance of a strut in the configurations from the subfamilies A, B and C considered above, reduces to taking the limit $p^2=q^2=1/2$; however, in no one case this limit leads to the equilibrium states with positive masses of both constituents. For instance, the equilibrium configuration due to Dietz and Hoenselaers \cite{DHo} which recently has been analyzed in the paper \cite{CMR} follows from (\ref{b_2}) as the limiting case
\be
p=q=\frac{1}{\sqrt{2}}, \quad \beta=\frac{1}{\sqrt{2}P-1}, \label{b_DH} \ee
and the corresponding total mass assumes the form
\be
M=-\frac{4\kappa}{\sqrt{2}P+3}. \label{M_DH} \ee
Hence $M$ is a negative quantity for all $\kappa>0$, $|P|\le 1$.

\section{Conclusion}

We have succeeded in obtaining a concise form of the general five-parameter K-C metric by making use of Perj\'es' representation of T-S solutions. As a convincing application of our results we have described analytically the general class of stationary axisymmetric configurations composed of two extreme Kerr sources separated by a massless strut. The existence of the black-hole sector in such configurations (characterized by positive Komar masses of both constituents) suggests that these binary systems can be regarded as carriers of interesting physical information about interacting extreme black holes, whose entire importance has yet to be clarified in the future.

\section*{Acknowledgements}
We would like to thank Erasmo G\'omez for technical support which was indispensable for successful completion of the complicated computer calculations. We are also thankful to the referee for a useful suggestion. This work was partially supported by Project FIS2006-05319 from Ministerio de Ciencia y Tecnolog\'\i a, Spain, and by the Junta de Castilla y Le\'on under the ``Programa de Financiaci\'on de la Actividad Investigadora del Grupo de Excelencia GR-234'', Spain.

\appendix
\section{The K-C solution in terms of determinants} %Empty argument \section{} yields `Appendix'.
%
%\section{Second Appendix}

The K-C solution is a vacuum specialization of the nine-parameter electrovac rational function solution considered in the paper \cite{MSM} and, therefore, its metric functions $f$, $\gamma$, $\omega$ are defined by the following expressions:\footnote{Note that $E$ and $\bar E$ in the numerator of the function $\omega$ from Eq.~(14) of Ref.~[29] must have the subscript `--'.}
\be f=\frac{E_+\bar E_-+\bar E_+E_-}{2E_-\bar E_-}, \quad e^{2\gamma}=\frac{E_+\bar E_-+\bar E_+E_-}{2K_0\bar K_0r_1^4r_2^4}, \quad \omega=-\frac{4{\rm Im}(\bar E_-G)}{E_+\bar E_-+\bar E_+E_-}, \label{A1} \ee
where $r_i=\sqrt{\rho^2+(z-\a_i)^2}$, the determinants $E_\pm$ and $G$ have the form
\bea
&&E_\pm=\left|\begin{array}{ccccc}
1 & \hspace{0.2cm}1 & \hspace{0.2cm}1 & (z-\a_1)/r_1 & (z-\a_2)/r_2 \vspace{0.1cm}\\
\pm1 &  & & & \vspace{0.2cm}\\ \pm1 & & & & \\\vspace{-0.35cm}  & &  & {\cal M} & \\
0 & & & & \vspace{0.2cm}\\ 0 & & & & \\
\end{array}\right|, \nonumber\\ &&G=\left|\begin{array}{ccccc}
0 & r_1+\a_1-z & r_2+\a_2-z & \rho^2/r_1 & \rho^2/r_2 \vspace{0.1cm} \\ -1 &  & & & \vspace{0.2cm} \\ -1 & & & & \\\vspace{-0.35cm}  & & {\cal M} &  & \\
0 & & & & \vspace{0.2cm}\\ 0 & & & & \\
\end{array}\right|, \label{A2}\\ \nonumber\\ &&{\cal M}=\left(\begin{array}{cccc} \dfrac{r_1}{\alpha_1 -
\beta_1} & \dfrac{r_2}{\alpha_2 - \beta_1} & - \dfrac{r_1^2}{(\alpha_1 - \beta_1)^2}
& - \dfrac{r_2^2}{(\alpha_2 - \beta_1)^2}\vspace{0.25cm}\\
\vspace{0.25cm}\dfrac{r_1}{\alpha_1 - \beta_2} &
\dfrac{r_2}{\alpha_2 - \beta_2} & - \dfrac{r_1^2}{(\alpha_1 - \beta_2)^2}
& - \dfrac{r_2^2}{(\alpha_2 - \beta_2)^2}\\
\vspace{0.25cm}\dfrac{1}{\alpha_1 - \bar\beta_1} &
\dfrac{1}{\alpha_2 - \bar\beta_1} & r_1^2\dfrac{\partial} {\partial
\alpha_1}\left[\dfrac{1}
{(\alpha_1-\bar\beta_1)r_1}\right] & r_2^2\dfrac{\partial} {\partial
\alpha_2}\left[\dfrac{1}
{(\alpha_2-\bar\beta_1)r_2}\right]\\
\dfrac{1}{\alpha_1 - \bar\beta_2} & \dfrac{1}{\alpha_2 -
\bar\beta_2} & r_1^2\dfrac{\partial} {\partial
\alpha_1}\left[\dfrac{1}
{(\alpha_1-\bar\beta_2)r_1}\right] &
r_2^2\dfrac{\partial} {\partial
\alpha_2}\left[\dfrac{1}
{(\alpha_2-\bar\beta_2)r_2}\right]\\
\end{array}\right), \nonumber  \eea
and the determinant $K_0$ is defined by the $4\times4$ matrix obtainable from ${\cal M}$ by simply setting $r_1$ and $r_2$ to unity.

Expressions (\ref{A2}) contain two arbitrary complex parameters $\beta_1$, $\beta_2$ and two real parameters $\a_1$, $\a_2$ which can be subjected to the constraint $\a_1+\a_2=0$. The parametrizations in (\ref{A2}) and in the original K-C paper \cite{KCh} are equivalent and related with each other by the formulas
\bea
&&\a_1=-\a_2=\kappa, \nonumber\\
&&\beta_1+\beta_2=-\frac{2\kappa(p+q\beta-ip\a) \exp(-i\gamma)-2i\kappa(pq+\beta)}{p^2+\a^2-\beta^2}, \nonumber\\ &&\beta_1 \beta_2=\frac{\kappa^2(p^2-\a^2+\beta^2)+2i\kappa^2[(p\beta+iq\a) \exp(-i\gamma)+\a]}{p^2+\a^2-\beta^2}. \label{A3} \eea

Formulas (\ref{A1}) and (\ref{A2}), after expanding the determinants, introducing the prolate spheroidal coordinates (\ref{xyrz}) and passing to the K-C parameters via (\ref{A3}), eventually yield Eqs.~(\ref{mf_gen}) of Section 3.

\end{document}